\documentclass[twocolumn,showpacs,preprintnumbers,superscriptaddress,amsmath,amssymb,prl]{revtex4}

\usepackage{amssymb}
\usepackage{stmaryrd}
\usepackage{amsmath}
\usepackage{txfonts}
\usepackage{graphicx}
\usepackage{CJK}

\usepackage{bm}
\usepackage{ulem}

\begin{document}
\begin{CJK*}{GB}{SongMT}
\CJKfamily{gbsn}

\title{Trajectory-Based Unveiling of Angular Momentum of Photons}

\author{Yongnan Li}
\affiliation{School of Physics and Key Laboratory of Weak-Light Nonlinear Photonics, Nankai University, Tianjin 300071, China}

\author{Zhi-Cheng Ren}
\affiliation{School of Physics and Key Laboratory of Weak-Light Nonlinear Photonics, Nankai University, Tianjin 300071, China}

\author{Ling-Jun Kong}
\affiliation{School of Physics and Key Laboratory of Weak-Light Nonlinear Photonics, Nankai University, Tianjin 300071, China}

\author{Chenghou Tu}
\affiliation{School of Physics and Key Laboratory of Weak-Light Nonlinear Photonics, Nankai University, Tianjin 300071, China}

\author{Hui-Tian Wang}
\email{htwang@nju.edu.cn/htwang@nankai.edu.cn}
\affiliation{School of Physics and Key Laboratory of Weak-Light Nonlinear Photonics, Nankai University, Tianjin 300071, China}
\affiliation{National Laboratory of Solid State Microstructures, Nanjing University, Nanjing 210093, China}
\affiliation{Collaborative Innovation Center of Advanced Microstructures, Nanjing 210093, China}

\date{Oct. 7, 2015}

\begin{abstract}
\noindent
The Heisenberg uncertainty principle suggests that it is impossible to determine the trajectory of a quantum particle in the same way as a classical particle. However, we may still yield insight into novel behavior of photons based on the average photon trajectories (APTs). Here we explore the APTs of optical fields carrying spin angular momentum (SAM) and orbital angular momentum (OAM) under the paraxial condition. We define the helicity and differential helicity for unveiling the three-dimensional spiral structures of the APTs of optical fields carrying the SAM and/or the OAM. We clarify the novel behaviors of the APTs caused by the SAM and OAM as well as the SAM-OAM coupling. The APT concept is also very helpful for profoundly understanding the trapped particle motion and has the potential to elucidate other physical systems.
\end{abstract}

\pacs{42.50.Tx, 42.25.Ja, 87.80.Cc}


\maketitle
\end{CJK*}
\newpage

\noindent Heisenberg's statement that ``The more precisely the position is determined, the less precisely the momentum is known in this instant, and vice versa" \cite{R01}, conveys the fact that there is a limit to the precision to which the position and momentum of a quantum particle can be known simultaneously; that is, the trajectory of a single quantum particle cannot be as precise as that of a classical particle. As the motion of a classical particle is governed by Newtonian mechanics, knowledge of the position and momentum allows the past, present, and future states of the particle to also be known. Although the trajectory of an individual quantum particle is difficult to define because any measurement of the position (momentum) irrevocably perturbs the momentum (position), we may still gain some information without appreciably perturbing the future evolution of the quantum system through a weak measurement and determine a precise mean value for the observable of interest by averaging over many weak measurements \cite{R02}. For instance, the average trajectories of single photons has been investigated in a double-slit interferometer \cite{R03}.

Besides the linear momentum, photons can carry the angular momentum (AM), which is classified into spin angular momentum (SAM) and orbital angular momentum (OAM) \cite{R04,R05,R06}: the SAM is always associated with the polarization (SAM of $+ \hbar $, $- \hbar $ and 0 per photon for the right-circularly, left-circularly and linearly polarized light, respectively and $\hbar$ is the reduced Planck constant) \cite{R04,R05,R06}, while the OAM is associated with a helical or twisted wavefront of $\exp (i m\phi )$ (OAM of $m \hbar$ per photon, where $m$ is the topological charge) \cite{R04,R05,R06,R07,R08,R09}. The photon AM has attracted considerable interest in various realms, in optical manipulation \cite{R10,R11,R12}, optical communication \cite{R13,R14,R15}, and quantum optics \cite{R16,R17,R18,R19}.

In optical tweezers experiments, the photon AM can be observed through the rotation of the trapped microscopic particles. The SAM causes a trapped particle to rotate about its own axis \cite{R20}, while the OAM induces an orbital motion of the trapped particles \cite{R21}. In particular, under the nonparaxial condition, a focused circularly polarized field could drive the orbital motion of the particles owing to the SAM-to-OAM conversion caused by the induced additional helical phase of the longitudinal field component \cite{R22,R23,R24}. A new class of photon OAM associated with the curl of polarization independent of phase has been predicted and demonstrated, which differs from the well-known OAM associated with the phase gradient independent of polarization in that this novel OAM can be carried by a radial-variant vector field with hybrid polarization states \cite{R25}. Although a quantum particle is not allowed to move along a definite path due to its nonlocalization, the APTs related to large-scale properties in the quantum system exhibit signatures of underlying the classical dynamics \cite{R26}. Here we devote to unveil the photon AM based on the APT concept, including the SAM and the OAM as well as the SAM-OAM coupling, under the paraxial condition.

Under the paraxial approximation, a scalar Laguerre-Gaussian (LG) field propagating along the $+z$ direction will have a transverse electric field component that can be written in a cylindrical coordinate system $(r, \phi, z)$ as
\noindent
\begin{subequations}
\begin{align}\label{01}
\textbf{E} _ \bot (r, \phi, z) & = u (r, \phi, z) [ (\cos \phi + i \sigma \sin \phi) \textbf{\^{e}} _r \nonumber \\
& \quad \quad \quad \quad + (- \sin \phi + i \sigma \cos \phi) \textbf{\^{e}} _\phi ],
\end{align}
\noindent with
\begin{align}
u (r, \phi, z) & \propto \frac{1}{w(z)} \left[ \frac{r}{w(z)} \right]^{|m|} \exp \left[ - \frac{r^2}{w^2 (z)} \right] \nonumber \\
& \quad \times \exp \left[ i \frac{ k r^2}{ 2 R(z)} + i k z + i \zeta (z) \right] \exp (i m \phi),
\end{align}
\end{subequations}

\noindent where $w^2 (z) = w_0^2 (1 + z^2 /z_0^2)$ is the field radius, $R(z) = ( z^2 + z_0^2)/z$ is the radius of curvature of the wavefront, $\zeta (z) = (|m|+1) \arctan (z/z_0)$ is the Gouy's phase, $ z_0 = k w_0^2/2$ is the Rayleigh range, $ w_0 $ is the waist radius of the fundamental Gaussian mode, and $m$ is the topological charge. $\textbf{\^{e}} _r$ and $\textbf{\^{e}} _\phi$ are two transverse unit vectors in the radial and azimuthal directions, respectively. $\sigma$ describes the polarization state of the LG field: $\sigma  = \pm 1$ for right- and left-handed circular polarization, $\sigma \in (0, +1)$ for right-handed and $\sigma \in ( - 1, 0)$ for left-handed elliptical polarization, and $\sigma = 0$ for linear polarization, respectively. For any LG field in Eq.~(1), the single photons carry the OAM of $m \hbar$ \cite{R04,R05,R06,R07,R08,R09} and the intensity pattern is always a doughnut shape unless $m = 0$. The radius $R_0 (z)$ of the brightest intensity ring is $R_0 (z) = \sqrt{|m|/2} w (z)$ in the plane with a distance of $z$ from the waist plane $z = 0$ (in particular, $R_{00} = R_0 (z)|_{z = 0} = \sqrt{|m|/2} w_0$ at the waist), which is independent of the SAM or the polarization. When $m = 0$, the LG field degenerates into the well-known Gaussian field and the corresponding doughnut-shaped intensity pattern becomes into a round Gaussian profile with $R_0 (z) \equiv 0$ because the phase singularity at the field centre has disappeared.

Since the wave vector is always normal to the wavefront, an LG field should have a longitudinal field component. Under the paraxial approximation, for any optical field, the global electric and magnetic fields are

\noindent
\begin{subequations}
\begin{align}\label{02}
 \textbf{E}  & = \textbf{E} _ \bot + E_z \textbf{\^{e}} _z = \textbf{E} _ \bot + i k ^{-1} (\nabla _\bot \cdot \textbf{E} _\bot)  \textbf{\^{e}} _z, \\
 \textbf{H} & = \textbf{H} _ \bot + H_z \textbf{\^{e}} _z \propto \textbf{\^{e}} _z \times \textbf{E} _ \bot + i k ^{-1} \nabla _\bot \cdot (\textbf{\^{e}} _z \times \textbf{E} _ \bot) \textbf{\^{e}} _z,
\end{align}
\end{subequations}

\noindent where $\textbf{\^{e}} _z $ is the longitudinal unit vector and $\nabla  _\bot $ is the transverse gradient operator. For an \textit{ideal} plane wave, implying that $w_0 \rightarrow \infty$ or $u(r,\phi,z)$ is space invariant in Eq.~(1), the longitudinal field components $E_z$ and $H_z$ are null due to $\nabla _\bot \cdot \textbf{E} _\bot = 0$.

In the Bialynicki-Birula hydrodynamical frame, the electromagnetic energy flows along streamlines described by \cite{R27}

\noindent
\begin{equation}\label{03}
\frac{\textrm{d} \textbf{R} }{\textrm{d} s } = \frac{1}{c} \frac{ \textbf{S} (\textbf{R})}{ U ( \textbf{R} )},
\end{equation}
where $c$ is the speed of light, $s$ labels the envelope across the space of the corresponding streamline, $\textrm{d} \textbf{R} = \textrm{d} r \textbf{\^{e}} _r + r \textrm{d} \phi \textbf{\^{e}}_\phi + \textrm{d} z\textbf{\^{e}} _z$, $U (\textbf{R})$ is the time-averaged electromagnetic energy density, and $\textbf{S} ( \textbf{R} )$ is the time-averaged Poynting energy flow vector

\noindent
\begin{equation}\label{04}
\textbf{S} (\textbf{R}) \propto \mathfrak{Re} [ \textbf{E} ^* (\textbf{R}) \times \textbf{H} (\textbf{R})],
\end{equation}

\noindent where $\mathfrak{Re} [ \text{   } ]$ extracts the real part of the complex quantity. The solutions of Eq.~(3), the position coordinate $\textbf{R}(r, \phi, z)$, give the streamlines or the electromagnetic energy flow lines, and describe also the APTs within a Bohmian-like reinterpretation of the Bialynicki-Birula hydrodynamical formulation \cite{R27}.

For a paraxial polarized LG vortex field, with Eqs.~(1), (2) and (4), we yield
\begin{align}\label{05}
\textbf{S} (\textbf{R}) & \propto (1 + \sigma ^2 ) \textbf{\^{e}}_z + \frac{r (1 + \sigma ^2 )}{R (z)} \textbf{\^{e}} _r \nonumber \\
 & \quad + \left[ \frac{m (1 + \sigma ^2 )}{ k r} - \frac{ 2 |m| \sigma }{ k r} + \frac{ 4 r \sigma }{ k w^2 (z) } \right] \textbf{\^{e}}_\phi,
\end{align}

\noindent Substituting Eq.~(5) into Eq.~(3) yields differential equations for the APTs

\noindent
\begin{subequations}
\begin{align}\label{06}
 \frac{\textrm{d} r }{\textrm{d} z } & = \frac{r}{ R(z) }, \\
 \frac{\textrm{d} \phi}{\textrm{d} z} & = \frac{ 4 \sigma }{ (1 + \sigma ^2 ) k w^2 (z) } - \frac{ 2 |m| \sigma }{ (1 + \sigma ^2 ) k r^2 } + \frac{m}{ k r^2 }.
\end{align}
\end{subequations}

\noindent With Eq.~(6a) and with the aid of the above expression of $R(z)$, the radial coordinate $r(z)$ of the APT can be solved by
\begin{equation}\label{07}
r(z) = r_0 ( 1 + z^2 /z_0^2 )^{1/2},
\end{equation}

\noindent where $ r_0 $ is the initial radial coordinate of the photon in the input plane $z = 0$. Substituting Eq.~(7) into Eq.~(6b) yields easily an analytical solution of the cumulative spiral angle $\phi (z)$ of the APT during the propagation over a distance $z$
\noindent
\begin{equation}\label{08}
\phi (z) = \left[ \frac{ 2\sigma }{ (1 + \sigma ^2 ) } - \frac{ |m| \sigma w^2_0}{ (1 + \sigma ^2 ) r^2_0} + \frac{m w^2_0}{2 r^2_0} \right] \arctan \left( \frac{z}{ z_0 } \right).
\end{equation}

\noindent Equation~(8) indicates that the spiral angle $\phi (z)$ of the APT originates from the contributions of three parts: the SAM (first term), SAM-OAM coupling (second term), and OAM (third term). To quantitatively characterize the spiral degree of the APT, we define a parameter \textit{helicity H} to represent the average change in $\phi (z)$ over $z$. The global helicity $H$ can also be divided into three parts, $H = \phi (z) / z = H_{SAM} + H_{SOC} + H_{OAM}$
\begin{subequations}\label{09}
\begin{align}
H_{SAM} & = \frac{ 2\sigma }{ (1 + \sigma ^2 ) z } \arctan \left( \frac{z}{ z_0 } \right), \\
H_{SOC} & = - \frac{ |m| \sigma w^2_0}{ (1 + \sigma ^2) r^2_0 z} \arctan \left( \frac{z}{ z_0 } \right), \\
H_{OAM} & = \frac{m w^2_0}{2 r^2_0 z} \arctan \left( \frac{z}{ z_0 } \right).
\end{align}
\end{subequations}

\noindent The helicity thus reveals the correlations of the APT with the SAM, OAM, and SAM-OAM coupling. The periodicity of $\phi $ in the azimuthal dimension means that an APT will exhibit a helical propagation path, much like the shape of a vine. We define a very important parameter again, \textit{differential helicity} $H'= \textrm{d} \phi / \textrm{d} z$, which represents the local change of $\phi (z)$ for the APT in the plane $z$. In a similar way to $H$, $H'$ can also be classified into the contributions of three parts as $H' = H'_{SAM} + H'_{SOC} + H'_{OAM}$

\begin{subequations}\label{10}
\begin{align}
H'_{SAM} & = \frac{4 \sigma }{ (1 + \sigma ^2 ) k w^2_0 (1 + z^2 / z^2_0) }, \\
H'_{SOC} & = - \frac{2 |m| \sigma }{ (1 + \sigma ^2 ) k r^2_0 (1 + z^2 / z^2_0)}, \\
H'_{OAM} & = \frac{m}{k r^2_0 (1 + z^2/ z_0^2)}.
\end{align}
\end{subequations}

\noindent In should be emphasized that the angular velocities of the APTs, caused by the global AM, the SAM, the SAM-OAM couping and the OAM, should be proportional to the differential helicities, $H'$, $H'_{SAM}$, $H'_{SOC}$, and $H'_{OAM}$, respectively.

To visually view, we calculate the three-dimensional (3D) structures of the APTs for LG fields under the paraxial approximation. In the following calculations (Figs.~1 and 2), the used parameters are $\lambda = 633 $ nm (wavelength) and $w_0 = 1$ mm, and the waist of any LG field is located at the input plane $z = 0$. With Eq.~(1b), the LG field at $z = 0$ should be
\begin{equation}\label{10}
u (r, \phi, 0) \propto \frac{1}{w_0} \left[ \frac{r}{w_0} \right]^{|m|} \exp \left[ - \frac{r^2}{w_0^2} \right] \exp (i m \phi).
\end{equation}

\begin{figure*}[bhpt]
  \centering{\includegraphics[width=12.5cm]{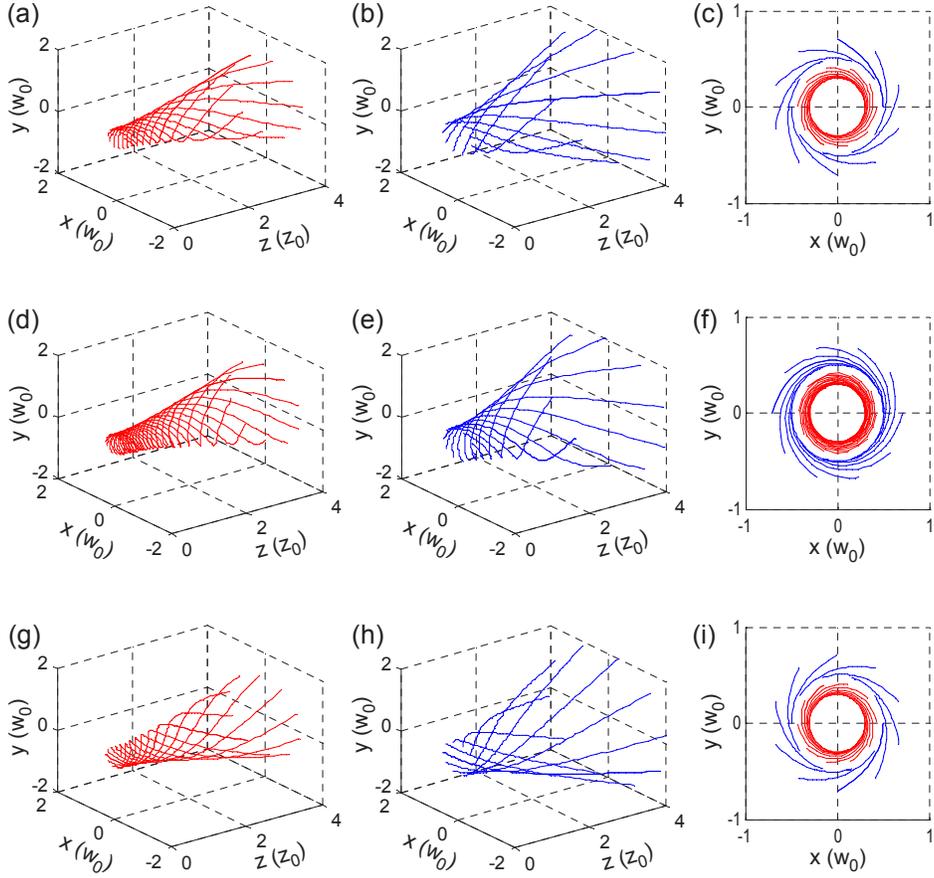}\\
  \caption{(color online) The 3D structures of the spiral APTs and their projections in the $\mathbf{\textit{z} = \textit{z}_0}$ plane for the linearly polarized LG vortex fields carrying different OAM. The waist of any linearly polarized $(\sigma = 0)$ LG vortex field is located in the input plane $z = 0$. First, second, and third rows show the spiral APTs for $m =1$, 2, and $-1$, respectively. Left and middle columns show the cases of $ r_0 = 0.3 w_0 $ and $ r_0 = 0.5 w_0 $, respectively. Right column shows the corresponding projections of the spiral APTs in the $z = z_0 $ plane.}}
\end{figure*}

Figure~1 depicts the 3D APTs for linearly polarized ($\sigma  = 0$, zero SAM) LG vortex fields (carrying the OAM of $m \hbar$). For $m = 1$ in Figs.~1(a)-(c) and $m = 2$ in Figs.~1(d)-(f), the LG fields with $m > 0$ have the right-handed spiral APTs, so the trapped microparticles will exhibit anticlockwise orbital motion \cite{R28}. For $m = -1$ in Figs. 1(g)-(i), however, the spiral APT of the LG field with $m < 0$ becomes left-handed, so the trapped microparticles will exhibit clockwise orbital motion in the opposite sense \cite{R28}. Compared Fig.~1(f) with Figs.~1(c) and (i), we see that for the projections of the APTs in the $z = z_0$ plane for $m = 2$ are denser than those for $m = \pm 1$. This suggests that the helicity of the spiral APT increases as $|m|$ enlarges. Therefore, the linearly polarized LG vortex fields with a higher OAM should result in the faster orbital motion of the trapped microparticles. These calculation results (Fig.~1) and the analytic expression in Eq.~(9a) show that for the LG vortex field carrying OAM only, the helicity of the spiral APT will decrease as the field propagates.

\begin{figure*}[htb]
  \centering{\includegraphics[width=12.5cm]{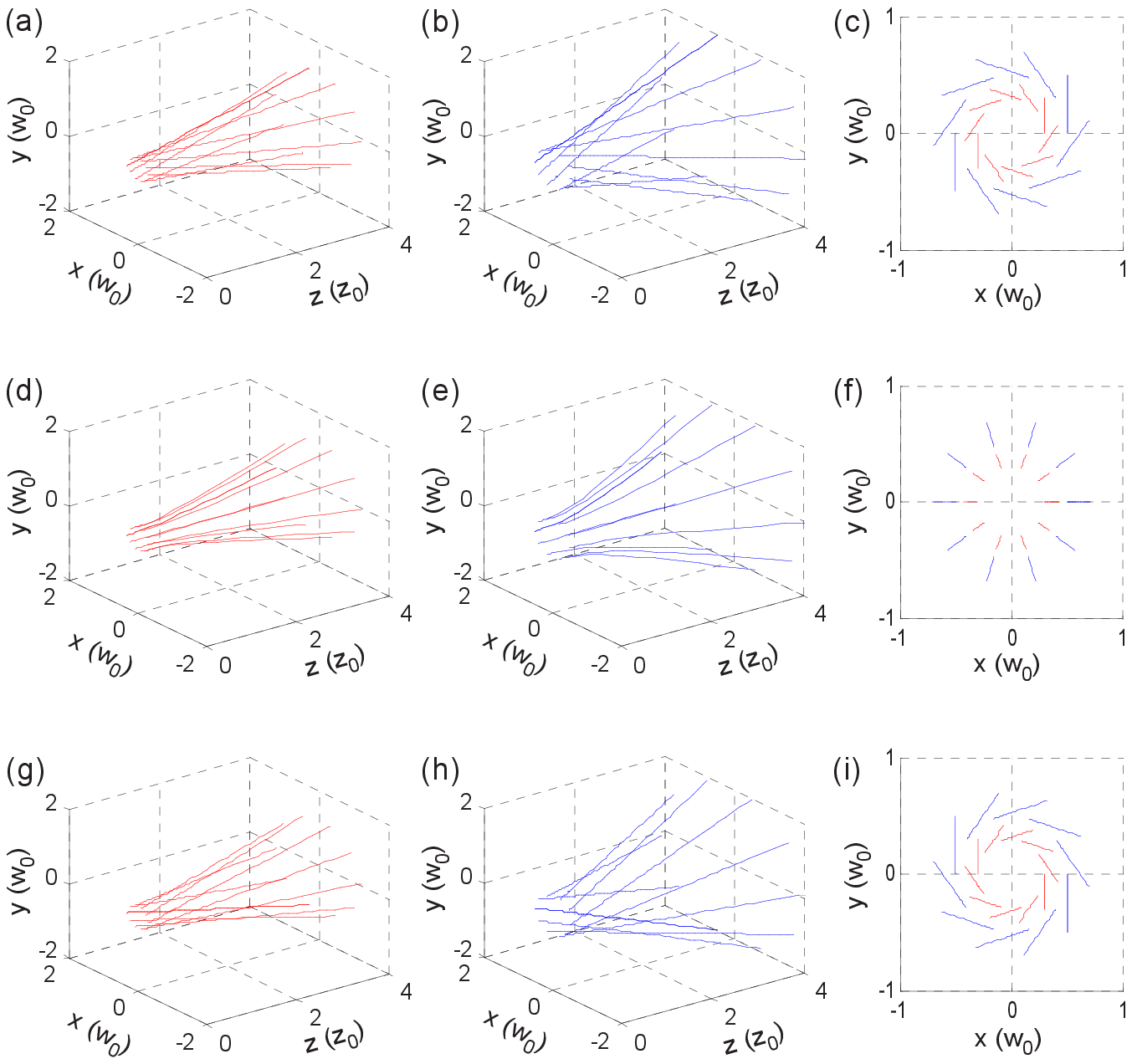}\\
  \caption{(color online) The 3D structures of the APTs and their projections in the $\mathbf{\textit{z} = \textit{z}_0}$ plane for the polarized Gaussian fields carrying no OAM. The waist of any polarized Gaussian field is located in the input plane $z = 0$. First, second, and third rows show the APTs for $\sigma = 1$, $0$, and $-1$, respectively. Left and middle columns correspond to the cases of $ r_0 = 0.3 w_0 $ and $ r_0 = 0.5 w_0 $, respectively. Right column shows the corresponding projections of the APTs in the $z = z_0$ plane.}}
\end{figure*}

Figure~2 shows the calculated 3D APTs for polarized LG fields carrying no OAM $(m = 0)$. In this case, the LG fields degenerate into polarized Gaussian fields. Clearly, the SAM, like the OAM, can also result in the spiral APTs with the same sense as the SAM. As shown in the second row of Fig.~2, for the linearly polarized field carrying no AM, the APTs do not exhibit a spiral structure. For circularly polarized fields with only the SAM, the helicity of the spiral APT will also decrease as the field propagates.

Figure~3 plots the dependence of the helicities of the spiral APTs caused by the SAM (or OAM) solely on the initial radial coordinate $r_0$ in the $z = z_0$ plane, for the different LG fields. For the circularly polarized Gaussian field (carrying the SAM solely) with a given waist $w_0$, the helicity of the spiral APT is independent of $r_0$, that is to say, the SAM-induced APTs have the same helicity regardless of the position. These results are in good agreement with the analytic expression in Eq.~(9a) and the 3D APTs shown in Fig.~2. For the linearly polarized LG fields carrying the OAM only, in contrast, the helicities of the APTs decrease linearly as $r^2_0$ increases. This means that photons far from the field axis have a smaller helicity, which is in agreement with the analytic expression in Eq.~(9c) and the 3D APTs shown in Fig.~1.

\begin{figure}[htbp]
  \centering{\includegraphics[width=8.5cm]{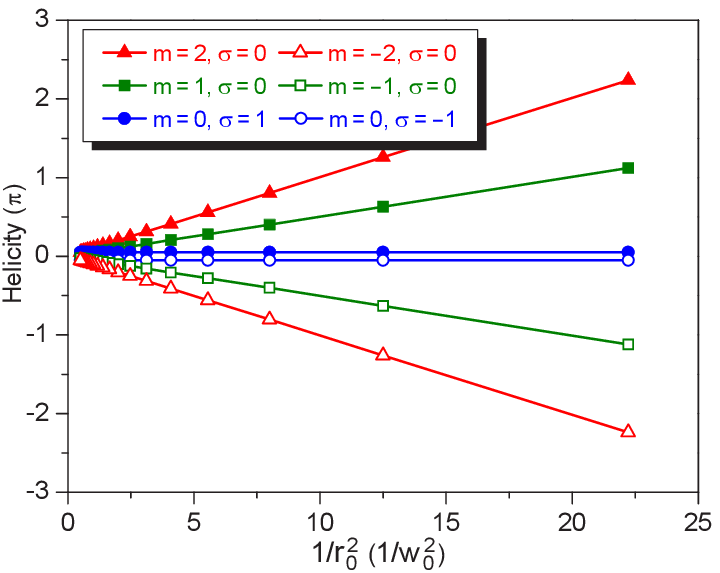}\\
  \caption{(color online) Dependence of the helicity of the APTs on the initial radial coordinate $r _0$ of photons in the $z = z _0$ plane. For the circularly polarized $(\sigma = \pm 1)$ Gaussian fields carrying no OAM $(m = 0)$ and the linearly-polarized $(\sigma = 0)$ LG vortex fields carrying the OAMs ($\pm 2 \hbar$ and $\pm \hbar$).}}
\end{figure}

In discussion, when the topological charge $m = 0$, the LG field described by Eq.~(1) becomes into a Gaussian field carrying no OAM. For a circularly polarized Gaussian field carrying the SAM of $\sigma \hbar$ only, we can find from Eqs.~(9a) and (10a) that the helicity $H_{SAM}$ and the differential helicity $H'_{SAM}$ of the APTs decrease as its waist radius $w_0$ increases. As is well known, the torque provided by the SAM of the circularly polarized Gaussian field can be transferred into a trapped birefringent particle to drive its rotation. The speed of rotation depends on the torque. The larger $H_{SAM}$ or $H'_{SAM}$ will provide the stronger torque. Hence the highly focused circularly polarized Gaussian field, which has the smaller waist $w_0$, contributes significantly to the rotation of the trapped particle. In an extreme situation when the circularly polarized Gaussian field degenerates into an \textit{ideal} circularly polarized plane wave field (implying that $w_0 \rightarrow \infty$ and $\nabla _\bot \cdot \textbf{E} _\bot \equiv 0$), its longitudinal field component will then be null. In this situation, although the ideal plane wave field carries an intrinsic SAM, the APTs do not exhibit a helical structure, and so no angular momentum would be transferred to a birefringent particle to cause it to rotate. Nevertheless, the particle rotation driven by the circularly polarized field has been indeed observed in the Beth's famous experiment \cite{R20}, which seems a paradox. However, this is in fact only a pseudo-paradox, as a circularly polarized ideal plane wave with infinite transverse dimensions does not exist. For any LG field (including a fundamental Gaussian field) that is a paraxial solution of the Maxwell's equations, its wavefront is in general a spherical surface excluding the waist plane, so the rotation of birefringent particles by a circularly polarized field is possible due to the presence of the spiral APTs.

For a circularly polarized Gaussian field carrying the SAM only and no OAM ($m = 0$), from the analytical expressions in Eqs.~(9) and (10) as well as the calculation results (Figs.~2 and 3), $H$ and $H'$ are independent of the radial position of the photon. This is very similar to the rotation of the Earth in that every location on the Earth has the same rotation angle and angular speed.

\begin{figure*}[bhtp]
  \centering{\includegraphics[width=12.5cm]{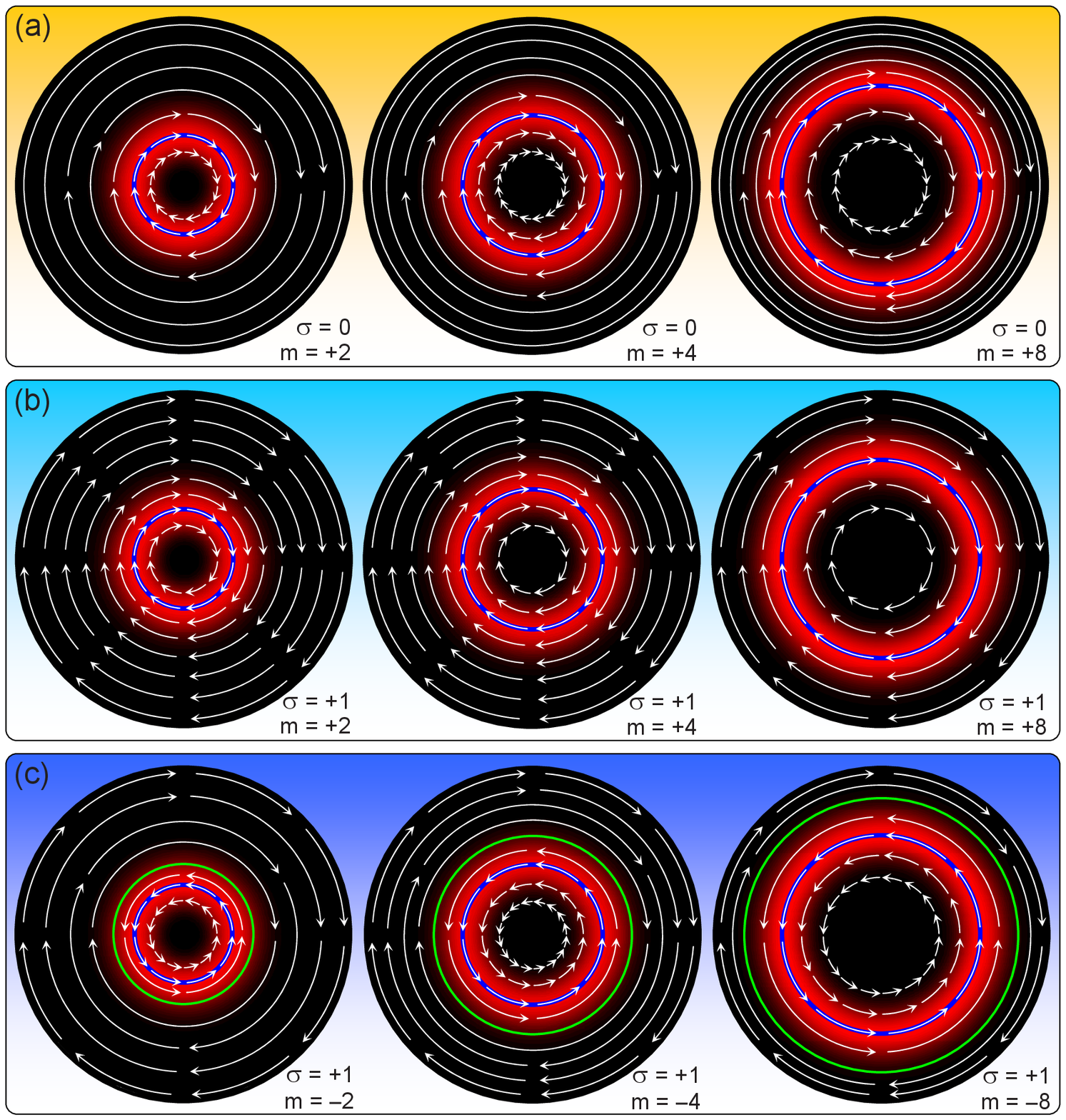}
  \caption{(color online) Schematic diagrams of the spatial distributions of the differential helicities of the APTs for three kinds of LG fields. (a) First kind of LG fields carrying the OAM solely, which include three zero-SAM ($\sigma = 0$) LG fields with different OAMs ($m = +2, +4, +8$). (b) Second kind of LG fields carrying the SAM and the OAM simultaneously, and the SAM and the OAM have the same senses, where $\sigma = + 1$ and $m = +2, +4, +8$. (c) Third kind of LG fields carrying the SAM and the OAM simultaneously, and the SAM and the OAM have the opposite senses, where $\sigma = + 1$ and $m = -2, -4, -8$. Any blue circle indicates the brightest ring of the donut-shaped intensity profile of the LG field. On any circle with a fixed radius, the direction of any arrow shows the direction of the local differential helicity of the APT and the number of arrows shows schematically the magnitude of the local differential helicity of the APT. Any dark green circle with no arrow in (c) indicates the local differential helicity there to be zero.}}
\end{figure*}

The radially-variant term $r^{|m|}$ in Eq.~(1b) plays a key role in the SAM-OAM coupling contribution to the helicity $H$ or differential helicity $H'$ of the APTs. For instance, if the LG field ($m \neq 0$) becomes a hypergeometric Gaussian field, that is, if there is no $r^{|m|}$ term in Eq.~(1b), then there will be no SAM-OAM coupling contribution to $H$ or $H'$ of the APTs in Eqs.~(9b) and (10b). Alternatively, if the $r^{|m|}$ term in Eq.~(1b) is modified instead of the power $|m|$ of $r$, the SAM-OAM coupling contribution to $H$ and $H'$ will not completely counterbalance the contribution from the OAM, even though the OAM has the same sense as the SAM. Therefore, the helicity caused by the SAM or the OAM is intrinsic, whereas the helicity originating from the SAM-OAM coupling is extrinsic.

If the LG field carries the OAM ($m \neq 0$) only and no SAM ($\sigma = 0$), from the analytical expressions in Eqs.~(6)-(10) and the calculation results in Figs.~1 and 3, $H$ and $H'$ of the spiral APTs caused by the OAM only depend on the radial position of the photon and decrease rapidly with distance from the field axis [Fig.~4(a)]. This is very similar to the tornado (a vortex of air), the wind speed decreases from the centre. Very interestingly, we find with Eq.~(10) that $H'$ of the spiral APTs caused by the OAM solely should be $H'|_{r(z) = R_0 (z)} = H'_{OAM}|_{r(z) = R_0 (z)} = 2 (m/|m|) k^{-1} w^{-2} (z)$ at the brightest ring [blue circle in Fig.~4(a)] with a radius of $R_0 (z) = \sqrt{|m|/2} w (z)$ in the plane $z$. Furthermore, $H'$ reaches its maximum, $H'|_{r_0 = R_{00}} = H'_{OAM}|_{r_0 = R_{00}} = 2 (m/|m|) k^{-1} w^{-2}_0$ ($\left\| H'|_{r_0 = R_{00}} \right\| = 2 k^{-1} w^{-2}_0$), at the brightest ring with a radius of $R_{00} = R_{0} (z) = \sqrt{|m|/2} w_0$ in the waist plane ($z = 0$) along the $z$ direction, and the spiral APTs has the same sense as the OAM.

If the LG field carries the SAM and the OAM with the \textit{same} sense, we can find from Eqs.~(6)-(10) that $H'_{SOC}$ counteracts completely $H'_{OAM}$ and $H'$ reduces then to $H'_{SAM}$. This gives rise to a very interesting phenomenon that, since any circularly polarized LG field carrying the OAM ($m \neq 0$) exhibits always a doughnut-shaped intensity pattern, the SAM will drive the orbit-like motion of the particles trapped in the brightest ring. Although the OAM has no direct contribution to the orbit-like motion of the trapped particles, the topological phase singularity or the OAM plays an indispensable role in the SAM-driven orbit-like motion or in the SAM-to-OAM-like conversion. In this case, the net $H'$ is equal to the SAM-induced $H'_{SAM}$ ($H' = H'_{SAM}$) is independent of the radial position [Fig.~4(b)]. Very interestingly, we find with Eq.~(10) that $H'$ of the spiral APT should be $H'|_{r(z) = R_0 (z)} = H'_{SAM}|_{r(z) = R_0 (z)} = 2 (\sigma/|\sigma|) k^{-1} w^{-2} (z)= 2 (m/|m|) k^{-1} w^{-2} (z)$ at the brightest ring [blue circle in Fig.~4(b)] with a radius of $R_0 (z) = \sqrt{|m|/2} w (z)$ in the plane $z$. Of course, $H'$ reaches also its maximum $H'|_{r_0 = R_{00}} = H'_{SAM}|_{r_0 = R_{00}} = 2 (\sigma/|\sigma|) k^{-1} w^{-2}_0 = 2 (m/|m|) k^{-1} w^{-2}_0$ ($\left\| H'|_{r_0 = R_{00}} \right\| = 2 k^{-1} w^{-2}_0$) at the brightest ring with a radius of $R_{00} = R_{0} (z) = \sqrt{|m|/2} w_0$ in the waist plane ($z = 0$) along the $z$ direction. The spiral APTs have the same sense as the SAM or the OAM.

If the SAM and the OAM have the \textit{opposite} sense, we discover another very interesting phenomenon. With Eqs.~(6)-(10), we find a \textit{boundary} being special radial position $r (z) = \widetilde{R}_{0} (z) = \sqrt{|m|} w (z)$ in the plane $z$ or in a special initial radial position $r_0 = \widetilde{R}_{00} = \sqrt{|m|} w_0$ in the waist plane $z = 0$ [green circle in Fig.~4(c)], at which the photons have net zero helicity $(H = 0)$ and net zero differential helicity $(H' = 0)$. This is quite different from the cases shown in Figs.~4(a)  and 4(b). Photons within the region $r (z) > \widetilde{R}_{0} (z)$ in the plane $z$ or $r_0 > \widetilde{R}_{00}$ in the waist plane $z = 0$ have spiral APTs dominated by the SAM (the same sense as the SAM), whereas those photons within the region $r (z) < \widetilde{R}_{0} (z)$ or $r_0 < \widetilde{R}_{00}$ are governed by the OAM (the same sense as the OAM), as shown in Fig.~4(c). As $r_0$ increases from $\widetilde{R}_{00}$ to $\infty$ , $H'$ increases from zero to $H'_{SAM}$ because the contribution of the OAM becomes null when $r_0$~$\rightarrow$~$\infty$. When $r_0$ is gradually decreased from $\widetilde{R}_{00}$, $H'$ will be gradually enlarged from zero. Most of the energy of the LG field is within the region $r (z) < \widetilde{R}_{0} (z)$ or $r_0 < \widetilde{R}_{00}$ because $\widetilde{R}_{0} (z) = \sqrt{|m|} w(z) > R_{0} (z) = \sqrt{|m|/2} w (z)$ or $\widetilde{R}_{00} = \sqrt{|m|} w_0 >R_{00} =\sqrt{|m|/2} w_0$ [Fig.~4(c)]. Interestingly, we also verify that photons located in the brightest ring [blue circle at $r(z) = R_{0} (z)$ or $r_0 = R_{00}$ in Fig.~4(c)] have $H'|_{r(z) = R_{0}(z)} = 2 (m/|m|) k^{-1} w^{-2} (z)$ or $H'|_{r_0 = R_{00}} = 2 (m/|m|) k^{-1} w^{-2}_0$, so that $H'|_{r(z) = R_{0}(z)}$ or $H'|_{r_0 = R_{00}}$ has the same magnitude as the SAM, but the opposite sense to the SAM and the same sense to the OAM.

As discussed above, we confirmed that as long as a LG field carries the OAM, regardless of whether it carries the SAM and whether the relative sense between the OAM and the SAM, the spiral APTs of the photons located at the brightest ring have always an identical differential helicity as $H'|_{r(z) = R_{0}(z)} = 2 (m/|m|) k^{-1} w^{-2}(z)$ or $H'|_{r_0 = R_{00}} = 2 (m/|m|) k^{-1} w^{-2}_0$. Clearly, its magnitude is independent of the OAM (the topological charge $m$) and its sense is always the same as the OAM. As a result, the angular velocity of the APTs of photons located in the brightest ring should be identical for the LG field carrying any OAM (any $m$); but the linear velocity of the APTs linearly increases as $\sqrt{|m|}$ because the radius of the brightest ring is in direct proportion to $\sqrt{|m|}$. The APTs concept should be an effective way for profoundly understanding the motion of the dielectric particles trapped by the LG fields. As is well known, the dielectric particles will be trapped in the brightest ring of the LG field. The orbital motion of the trapped particles along the brightest ring will depend on two factors: the local maximum intensity and the local differential helicity of the APTs. A nonzero local differential helicity is a necessary condition for providing the torque, and a stronger local intensity is beneficial for driving the orbital motion. As discussed above, although the radius of the brightest ring increases as $|m|$ or the OAM increases, $\left\| H'|_{r_0 = R_{00}} \right\|$ is independent of $|m|$. Under the assumption that the viscous resistance can be ignored, the angular velocity of the orbital motion should be independent of $m$, implying that the increase of $|m|$ cannot raise the angular velocity. However, the linear velocity of the orbital motion increases as $|m|$ or the OAM increases because the radius of the brightest ring is directly proportional to $\sqrt{|m|}$. Therefore, the motion is faster when the LG field carries a higher OAM (or a larger $|m|$). It should be emphasized that the most efficient driving of the motion of the trapped particles occurs when the particles are trapped at the waist of the LG field because there the local light intensity and the local $H'$ are both maximum.

In conclusion, although it is impossible to rigorously discuss the trajectory of an individual quantum particle, we can obtain the ``average photon trajectories." We define the helicity and differential helicity for unveiling the 3D spiral structures of the APTs of the LG fields carrying the SAM and/or the OAM. The APT concept is of great importance for exploring the fundamental insights into the nature of light and offers an alternative route for unveiling the angular momentum of light and for profoundly understanding the motion of the trapped particles in tweezers experiments. In addition, the electrons can be accelerated continuously along the circularly-polarized laser propagation direction and the collimated relativistic electron beams have the 3D spiral structure \cite{R29}, which indirectly indicates the spiral APTs of the photons. The average trajectories can be also applied to understand the AM of vortex electron beams \cite{R30,R31} and the spin-to-orbit interaction processes \cite{R32}.

This work is supported by the 973 Program of China under Grant No. 2012CB921900, the National Natural Science Foundation of China under Grant No. 11534006, and the National scientific instrument and equipment development project 2012YQ17004, and the National Natural Science Foundation of China under Grants 11274183 and 11374166.


\end{document}